\documentclass{article}
\usepackage{spconf,amsmath,graphicx}
\usepackage{booktabs}
\usepackage{float}
\usepackage{cite}
\usepackage[T1]{fontenc}
\usepackage{amsfonts}
\usepackage{bm}
\usepackage{algorithm}
\usepackage[noend]{algpseudocode}
\usepackage{amsmath}
\usepackage{amsfonts}
\usepackage{multirow}

\title{harmonic gated compensation network plus for ICASSP 2022 DNS Challenge}
\name{Tianrui Wang$^{\dagger \star}$, Weibin Zhu$^{\dagger}$, Yingying Gao$^{\ddagger}$, Yanan Chen$^{\ddagger}$, Junlan Feng$^{\ddagger}$, Shilei Zhang$^{\ddagger}$}
\address{$^{\dagger}$ Institute of Information Science, Beijing Jiaotong University, Beijing, China \\
	$^{\ddagger}$ China Mobile Research Institute, Beijing, China}

\begin{document}
	%
	\maketitle
	\begin{abstract}
		The harmonic structure of speech is resistant to noise, but the harmonics may still be partially masked by noise. Therefore, we previously proposed a harmonic gated compensation network (HGCN) to predict the full harmonic locations based on the unmasked harmonics and process the result of a coarse enhancement module to recover the masked harmonics. In addition, the auditory loudness loss function is used to train the network. For the DNS Challenge, we update HGCN with the following aspects, resulting in HGCN+. First, a high-band module is employed to help the model handle full-band signals. Second, cosine is used to model the harmonic structure more accurately. Then, the dual-path encoder and dual-path rnn (DPRNN) are introduced to take full advantage of the features. Finally, a gated residual linear structure replaces the gated convolution in the compensation module to increase the receptive field of frequency. The experimental results show that each updated module brings performance improvement to the model. HGCN+ also outperforms the referenced models on both wide-band and full-band test sets.
	\end{abstract}

	\section{Introduction}
	\label{sec:intro}
	\renewcommand{\thefootnote}{}{\footnote{$^\star$Work done during internship at China Mobile Research Institute.}}Speech enhancement (SE) aims to improve speech quality. Many researchers introduce intuitive signal processing ideas to deep learning \cite{crn,nsnet}. \cite{rnnoise} explores an SE model combining signal processing and deep learning, but simple structure limits its performance. \cite{dccrn} introduces the complex operations, which improve the performance but with less auditory characteristic. \cite{sdccrn,deepfilter} introduce the auditory feature, but spectra with wide frequency bandwidth degrade the performance. \cite{gaze} verifies the effectiveness of the hearing pipeline structure. \cite{phasen} verifies the necessity of modeling the harmonic. Inspired by these works, we proposed the HGCN for SE \cite{hgcn}. In HGCN, a high-resolution harmonic integration algorithm is proposed to predict the harmonic locations. Then the locations are used as a gate to help the subsequent module compensate for the result of the coarse enhancement module to obtain a refined result. To make the enhancement more consistent with human hearing, we previously proposed a loss function based on auditory loudness power compression (APC-SNR) \cite{apcsnr}. Both HGCN and APC-SNR have proved their effectiveness for SE.

	Compared to the HGCN, each module of the HGCN+ is updated and the HGCN+ can handle full-band (FB, 0\textasciitilde 24 KHz) signals. Since the wide-band (WB, 0\textasciitilde 8 KHz) is more likely to contain high energies, tonalities and long sustained sounds, while the high-band (HB, 8\textasciitilde 24 KHz) tends to have low energies, noise and rapidly decaying sounds \cite{wbandhb}, the HB and WB spectra are modeled separately \cite{sdccrn}. The HB spectrum is enhanced by a lightweight NSNet \cite{nsnet} and the WB spectrum is enhanced by an HGCN that is updated in the following aspects. 1) The dual-path encoder and DPRNN \cite{dpcrn,dprnn} are introduced to take full advantage of the features. 2) Cosine is adopted to model the harmonic peak-valley structure, and the voiced region detection (VRD) is judged based on the harmonic integration significance. 3) The gated convolution is replaced by a residual gated structure comprised of linear layers and Gated Recurrent Units (GRUs) \cite{GRUs} to increase the receptive field of frequency. Since we model the WB and HB spectra separately, HGCN+ can h andle both WB and FB signals without resampling. Experimental results show that HGCN+ outperforms the referenced methods on test sets.
	
	\begin{figure*}[htb]
		\centering
		\vspace{-0.3cm}
		\includegraphics[width=18cm]{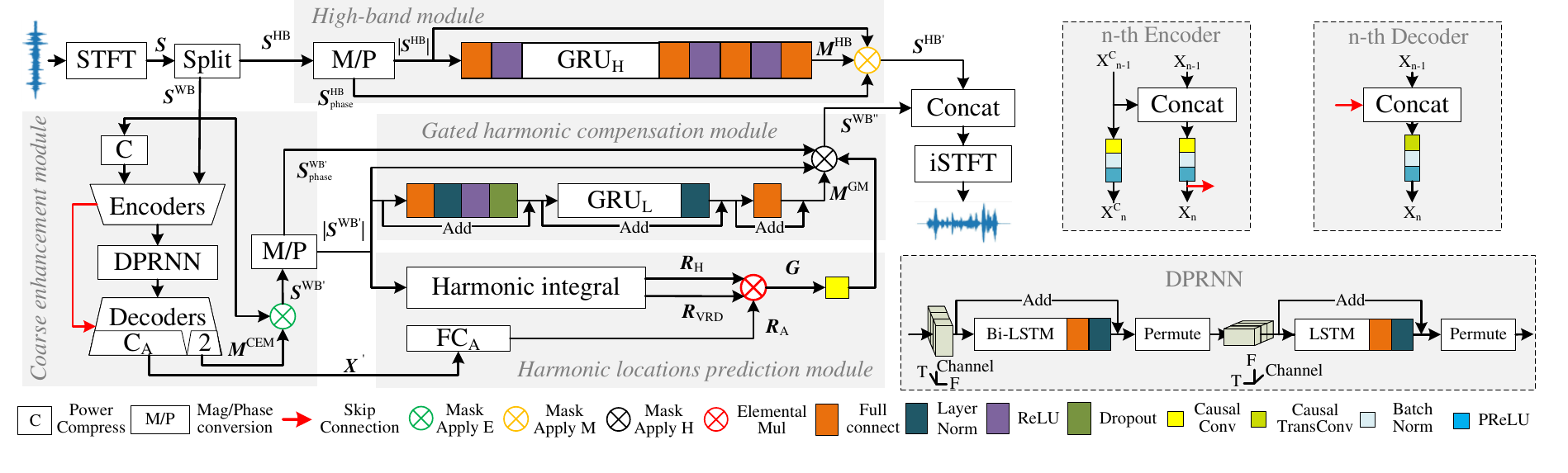}
		\vspace{-1.2cm}
		\caption{Architecture of the proposed HGCN+.}
		\vspace{-0.65cm}
		\label{HGCN}
	\end{figure*}
	\section{PROPOSED HGCN+}
	\label{sec:model}	
	
	The overall diagram of the HGCN+ is as shown in Fig.~\ref{HGCN}, which is comprised of four parts, namely the high-band module (HBM), coarse enhancement module (CEM), harmonic locations prediction module (HM), and gated harmonic compensation module (GHCM). The noisy signal is split into WB and HB spectra after a short-time Fourier transform (STFT). The HBM enhances the HB spectrum. And the WB spectrum is firstly passed to CEM to obtain a coarse result. Subsequently, HM predicts harmonic locations based on the coarse result. Then, GHCM compensates for the coarse result based on the harmonic location gates to get the refined WB result. Finally, the enhanced WB and HB spectra are concatenated and converted to waveform by an inverse STFT (iSTFT).
	
	\subsection{High-band model}
	HB spectra contain less speech information, and elaborate operation on them could bring a computational burden, so a lightweight magnitude module \cite{nsnet} is used. The HB spectrum $\bm{S}^\text{HB}=\text{Cat}(\bm{S}^{\text{HB}}_{r},\bm{S}^{\text{HB}}_{i}) \in \mathbb{R}^{T\times 2F}$ is firstly converted to $|\bm{S}^\text{HB}|$ and $\bm{S}^\text{HB}_\text{phase}$, where $|\bm{S}^\text{HB}|=({\bm{S}^{\text{HB}~2}_r}+{\bm{S}^\text{\text{HB}~2}_i})^{0.5}$ and ${\bm{S}^\text{HB}_\text{phase}}=\text{arctan}({\bm{S}^\text{HB}_i},{\bm{S}^\text{HB}_r})$ are the magnitude and phase, $\bm{S}^\text{HB}_r$ and $\bm{S}^\text{HB}_i$ are the real and imaginary parts, $T$ and $F$ denote the number of frames and the STFT bins respectively. The magnitude is used as input to predict the mask $\bm{M}^\text{HB}$. And HBM consists of fully connecting (FC), GRUs and ReLU \cite{relu}, as shown in Fig.~\ref{HGCN}. Finally, the HB enhanced result $\bm{S}^{\text{HB}'}$ is obtained by the Mask Apply M as follow,
	\begin{equation}
		\setlength{\abovedisplayskip}{2pt}
		\setlength{\belowdisplayskip}{2pt}
		\bm{S}^{\text{HB}'} = |\bm{S}^\text{HB}| \odot \sigma(\bm{M}^\text{HB}) \odot e^{j\bm{S}^\text{HB}_{\text{phase}}}
	\end{equation}
	where $\odot$ is the element-wise multiplication, $\sigma(\cdot)$ is sigmoid.

	\subsection{Coarse enhancement module}
	The noise with higher energy than speech is catastrophic for HM. So CEM is designed to roughly suppress the noise of WB spectra, which is an encoder-decoder structure. Before being input to the encoder, the WB spectrum $\bm{S}^{\text{WB}}=\text{Cat}(\bm{S}^{\text{WB}}_{r},\bm{S}^{\text{WB}}_{i})$ is compressed by a power of 0.23 and is input to the dual-path encoder together with the original spectrum, as shown in Fig.~\ref{HGCN}. Both the encoder and decoder are comprised of 2D causal convolution, batch normalization \cite{batchnorm}, and PReLU \cite{prelu}. Between the encoder and decoder, DPRNN \cite{dprnn,dpcrn} is inserted to model the multidimensional dependencies and Skip Connection concatenates the output of each encoder to the input of the corresponding decoder (red line in Fig.~\ref{HGCN}). The output channel of the last decoder is $(\text{C}_\text{A}+2)$, where 2 is the estimated mask of CEM $\bm{M}^{\text{CEM}}=\text{Cat}(\bm{M}^\text{CEM}_{r},\bm{M}^\text{CEM}_{i})$, $\text{C}_\text{A}$ is introduced in the next section. The Mask Apply E of CEM is as follows, 
	\begin{equation}
		\setlength{\abovedisplayskip}{2pt}
		\setlength{\belowdisplayskip}{0pt}
		\bm{S}^{\text{WB}'} = |\bm{S}^{\text{WB}}| \odot \text{tanh}(|\bm{M}^{\text{CEM}}|) \odot e^{j(\bm{S}^\text{WB}_{\text{phase}}+\bm{M}^\text{CEM}_{\text{phase}})}
		\label{maskapplyE}
	\end{equation}
	\subsection{Harmonic locations prediction module}
	\label{sec:harmonic}
\begin{figure*}[htb]
		\centering
		\includegraphics[width=18cm]{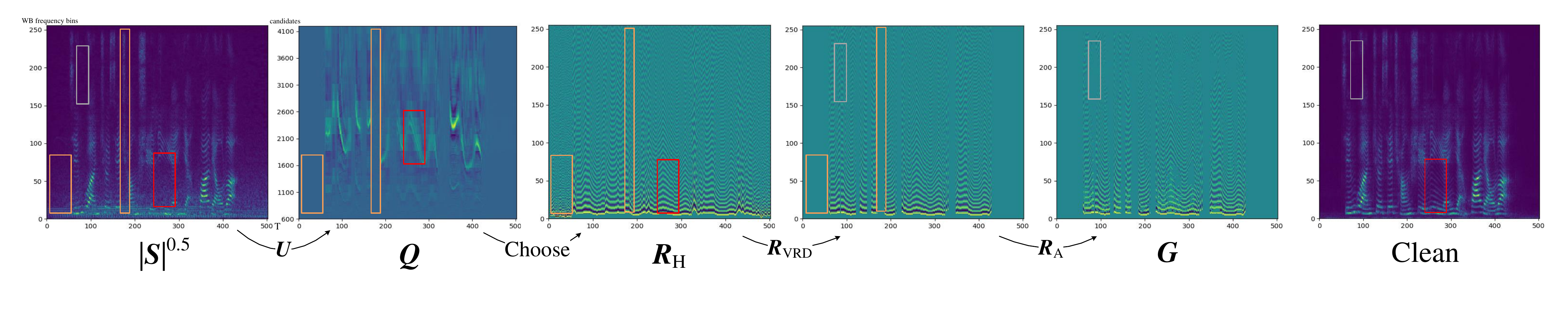}
		\vspace{-1.4cm}
		\caption{The calculation process of harmonic gate.}
		\label{harmonic}
		\vspace{-0.7cm}
	\end{figure*}

	The harmonics that are masked by noise can be deduced from the unmasked harmonics (red box in Fig.~\ref{harmonic}). \cite{swipe,HPS} proposed to model the peak-valley structure of the harmonics on spectra to detect the pitch. The pitch candidates are set first, and the integral of the multiple locations is taken as the significance $Q_{t,f_c}$ of each candidate $f_c$,
	\begin{equation}
		\setlength{\abovedisplayskip}{2pt}
		\setlength{\belowdisplayskip}{2pt}
		\label{si1}
		Q_{t,f_c}=\sum_{k=1}^{\text{8000}/f_c}(\frac{1}{\sqrt{k}}\cdot {|S^{\text{WB}'}_{t,kf_c}|}^{0.5}-\frac{1}{\sqrt{k}}{|S^{\text{WB}'}_{t,(k-\frac{1}{2})f_c}}|^{0.5})
	\end{equation}
	where $k$ denotes the multiple of the pitch. The candidate with the highest significance is regarded as the pitch.
	
\begin{figure}[htb]
		\vspace{-0.55cm}  
		\begin{minipage}[b]{.48\linewidth}
			\centering
			\centerline{\includegraphics[width=3.75cm]{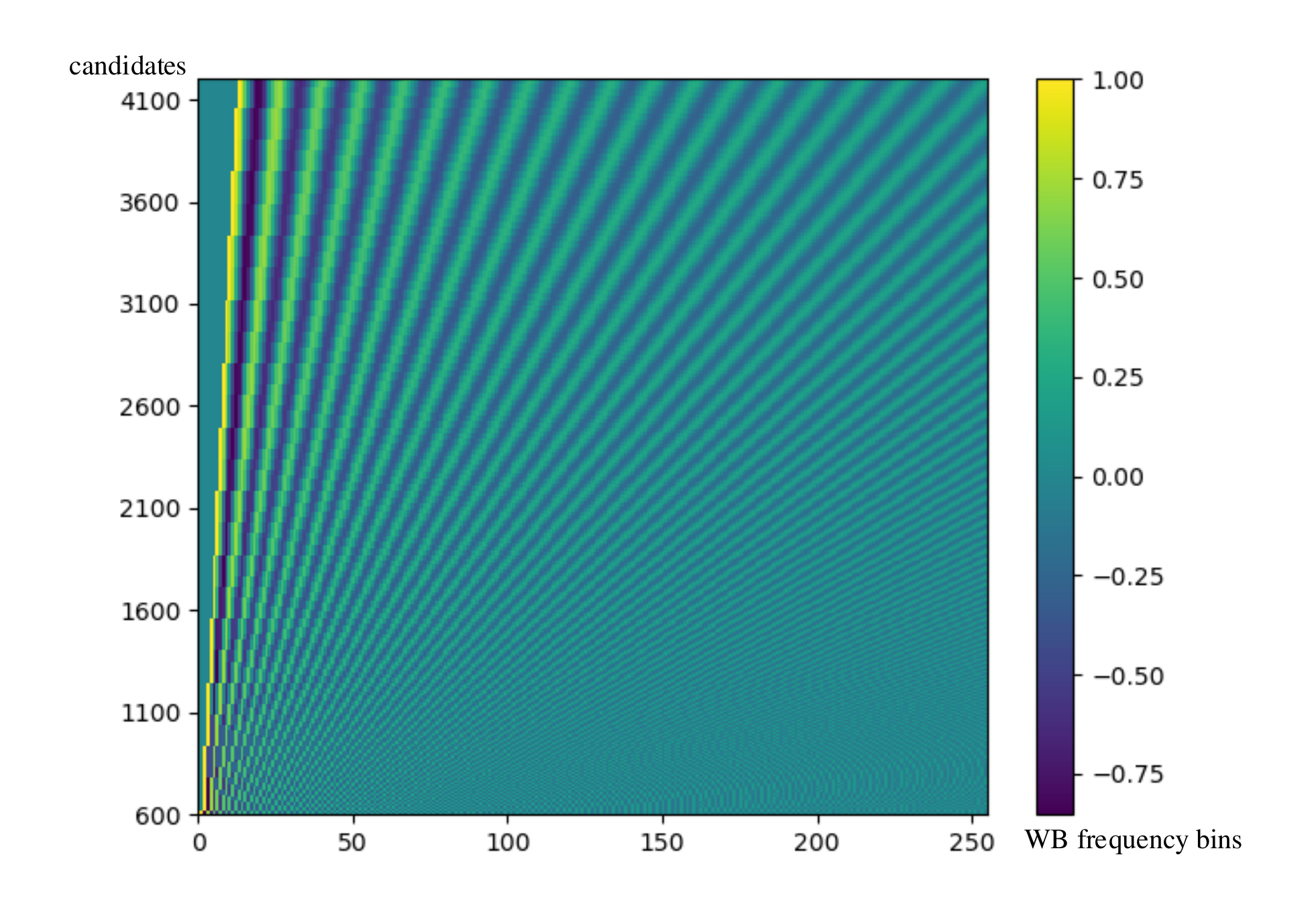}}
			\vspace{-0.3cm}
			\centerline{(a)}\medskip
		\end{minipage}
		\hfill
		\begin{minipage}[b]{0.48\linewidth}
			\centering
			\centerline{\includegraphics[width=3.5cm]{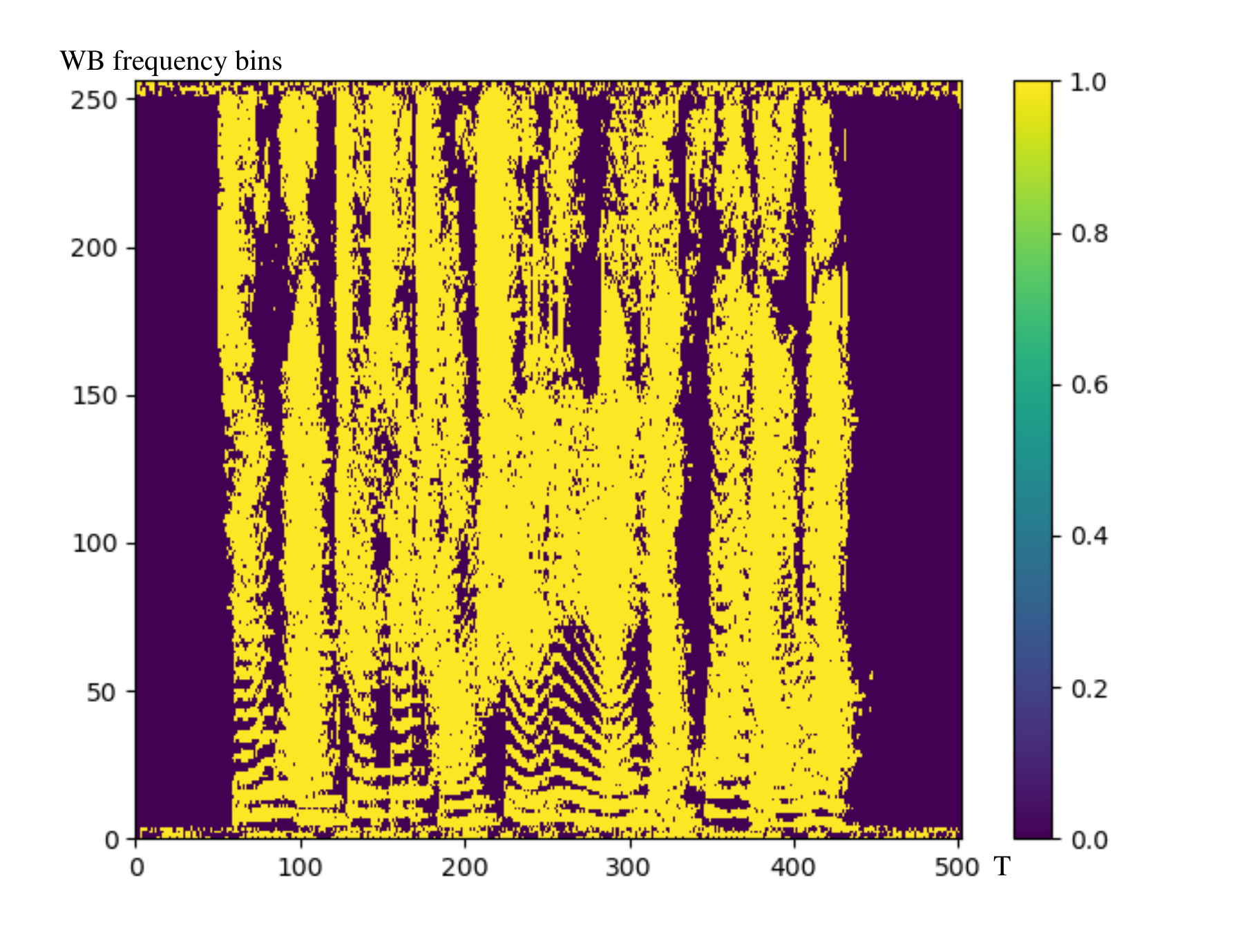}}
			\vspace{-0.3cm}
			\centerline{(b)}\medskip
		\end{minipage}
		\vspace{-0.7cm}
		\caption{(a) High-resolution harmonic integral matrix $\bm{U}$. (b) Labels for speech energy detection.}
		\vspace{-0.7cm}
		\label{UandLabel}
	\end{figure}

	To detect the pitch with fine-resolution based on STFT bins with wide-bandwidth, we proposed a high-resolution harmonic integration matrix $\bm{U}$ in HGCN, which sets candidates in 60\textasciitilde 420~Hz (normal pitch range of speech) with a resolution of 0.1~Hz. In this paper, the peak-valley modeling is improved by cosine function, and the $\bm{U}$ is designed as Algorithm~\ref{integralspectrum} and Fig.~\ref{UandLabel}(a), where $[\cdot]$ is a rounding operation, $\text{linspace}(a,b,c)$ generates an arithmetic progression between $a$ and $b$ of length $c$. Then the Eq.~(\ref{si1}) is updated to, 
	\begin{equation}
		\setlength{\abovedisplayskip}{3pt}
		\setlength{\belowdisplayskip}{3pt}
		\label{Q}
		\bm{Q}_t = |\bm{S}^{\text{WB}'}_t|^{0.5} \cdot \bm{U}^\top
	\end{equation}
	where $\bm{Q}_t \in \mathbb{R}^{1\times 3600}$ denotes the pitch candidate significances of the $t$-th frame. The candidate corresponding to the maximum value in $\bm{Q}_t$ is regarded as the pitch, and then the corresponding harmonic peak-valley structure is deduced based on the pitch as $\bm{R}_{\text{H}} \in \mathbb{R}^{T\times F}$, as shown in Fig.~\ref{harmonic}.
	
	There is no harmonic in unvoiced and silent frames (orange box in Fig.~\ref{harmonic}), so we apply the voiced region detector (VRD) to filter $\bm{R}_{\text{H}}$. In addition, the energy is low even if it's harmonic (gray box in Fig.~\ref{harmonic}), which needs to be filtered out. Therefore, the final harmonic gate $\bm{G}$ is calculated as follows,
	\begin{equation}
		\setlength{\abovedisplayskip}{3pt}
		\setlength{\belowdisplayskip}{3pt}
		\label{gate}
		\bm{G} = \bm{R}_{\text{VRD}} \odot \bm{R}_{\text{A}} \odot \bm{R}_{\text{H}}
	\end{equation}
	where $\bm{R}_{\text{A}} \in \mathbb{R}^{T\times F}$ denotes the non-low energy locations of speech and is detected by a speech energy detector (SED). Since SED needs to resist noise, we change the output channel number of the last CEM decoder to $(2+\text{C}_\text{A})$, and $\text{C}_\text{A}$ is the channels number of the input $\bm{X}^{'} \in \mathbb{R}^{T\times F \times \text{C}_{\text{A}}}$ for an FC layer ($\text{FC}_\text{A}$) to get a 2-D (low-high) classification probabilities $P_{t,f}=[p_0,p_1]$ for every T-F point $\bm{P} \in \mathbb{R}^{T\times F\times 2}$. And the $\bm{R}_{\text{A}}$ is obtained by $\bm{R}_{t,f} = \text{argmax}(P_{t,f})$. 
	
	The SED is designed to filter out the low energy parts, so we generate labels for SED based on energy. The mean of each bin in the clean logarithmic magnitude is counted as $\bm{\mu} = \sum_{t=1}^{T}\log{|\dot{\bm{S}}_{t}|}/T$, where $|\dot{\bm{S}}|$ represents the clean magnitude. And the label is 1 if the logarithmic magnitude of clean is larger than $\bm{\mu} \in \mathbb{R}^{F \times 1}$, 0 otherwise, as shown in Fig.~\ref{UandLabel}(b).

	The significances of the voiced frames are higher than that of the unvoiced and silent frames in the integral spectrum $\bm{Q}$, as shown in Fig.~\ref{harmonic}. So the VRD is designed as,
	\begin{equation}
		\setlength{\abovedisplayskip}{3pt}
		\setlength{\belowdisplayskip}{3pt}
		\label{voiced}
		(\bm{R}_\text{VRD})_{t} = \textit{\uppercase\expandafter{\romannumeral1}}(\max{(\bm{Q}_t)} > \left(\alpha \cdot \xi \right))     
	\end{equation}
	where $\max({\cdot})$ denotes the maximum value in the vector. If the input is true, $\textit{\uppercase\expandafter{\romannumeral1}}(\cdot)$ outputs 1. $\xi$ is the moving average which is updated as $\xi_{\text{new}} = 0.9\xi_{\text{old}} + 0.1 \sum_{t=1}^{T}\max{(\bm{Q}_t)}/T$. $\alpha$ is the scale factor ($\alpha$ is 0.4 in our experiments).
	
	\subsection{Gated harmonic compensation module}
	To recover the masked harmonics with the help of the unmasked harmonics, the module needs to have a wide receptive field of frequency, so a gated residual structure consisting of FCs and GRUs is adopted to compensate the coarsely processed spectra according to the harmonic locations $\bm{G}$, as shown in Fig.~\ref{HGCN}. And the harmonic gated compensation mask applying (Mask Apply H) is used as,
	\begin{equation}
		\setlength{\abovedisplayskip}{3pt}
		\setlength{\belowdisplayskip}{3pt}
		\begin{split}
			\bm{S}^{\text{WB}''} &=  \left[1+\text{CC}(\bm{G}) \odot \sigma(\bm{M}^{\text{GM}})\right] \odot |\bm{S}^{\text{WB}'}| \odot e^{j\bm{S}_{\text{phase}}^{\text{WB}'}}
		\end{split}
	\end{equation}
	where $\bm{M}^{\text{GM}}$ is the estimated mask of GHCM. Since some harmonic peaks maybe masked by noise, the magnitude that needs to be compensated is not a complete harmonic structure, a causal convolution ($\text{CC}$) is used to process the gate. 
	
	Finally, $\bm{S}^{\text{WB}''}$ and $\bm{S}^{\text{HB}'}$ are concatenated into FB complex spectra and converted to the waveform by iSTFT.
	
	\begin{algorithm}
		\centering
		\caption{Integral matrix}
		\label{integralspectrum}
		\begin{algorithmic}[1]
			\State $\bm{U} \gets \bm{0} \in \mathbb{R}^{3600 \times F}$
			\For{$f_c \gets 600 \to 4200$}
			\State $\text{loc}_\text{last} \gets 0$ ; $\text{peak}_\text{last} \gets 1$ ; $j \gets f_c-600$
			\For{$k\gets 1 \to \left[8000/(0.1\cdot f_c)\right]$}
			\State $\text{loc}\gets \left[0.1 \cdot f_c\cdot k\cdot F / 8000\right]$
			\State $\text{peak} \gets 1/\sqrt{k}$ ; $\bm{U}_{j,\text{loc}}\gets \text{peak}$
			\If{$\text{loc}-\text{loc}_\text{last}>1$}
			\State $\text{num}_\text{iner}=\text{loc}-\text{loc}_\text{last}$
			\State $\bm{F^\text{cos}} \gets \cos{(\text{linspace}(0,2\pi,\text{num}_\text{iner}))}$
			\State $\bm{F} \gets \text{linspace}(\text{peak}_\text{last},\text{peak},\text{num}_\text{iner})$
			\For{$i \gets 1 \to \text{num}_\text{iner}$}	
			\State $\bm{U}_{j,i+\text{loc}_\text{last}}=\bm{F^\text{cos}}_i \cdot \bm{F}_i$
			\EndFor		
			\Else 
			\State $\bm{U}_{j,\text{loc}} \gets \bm{U}_{j,\text{loc}}-(\text{peak}_\text{last}+\text{peak})/2$
			\State $\bm{U}_{j,\text{loc}_\text{last}}\gets \bm{U}_{j,\text{loc}_\text{last}}-(\text{peak}_\text{last}+\text{peak})/2$
			\EndIf
			\State $\text{loc}_\text{last} \gets \text{loc}$ ; $\text{peak}_\text{last} \gets \text{peak}$
			\EndFor
			\EndFor
		\end{algorithmic}
	\end{algorithm}

	\subsection{Loss function}
	The magnitude loss is used to model the HB magnitude. To make the  WB enhancement results consistent with the human hearing, we introduce the time-domain SI-SNR to measure the complex spectrum compressed by the auditory loudness power exponent $\bm{\gamma}$ \cite{apcsnr}. In addition, FocalLoss \cite{focalloss} is used as a loss function for the SED. The loss function of the whole model is defined as,
	\begin{equation}
		\setlength{\abovedisplayskip}{3pt}
		\setlength{\belowdisplayskip}{3pt}
		L = L_\text{HB} + L_\text{APC}^{\bm{S}^{\text{WB}'}} + L_\text{APC}^{\bm{S}^{\text{WB}''}}  + L_\text{focal}
	\end{equation}
	\begin{equation}
		\setlength{\abovedisplayskip}{1pt}
		\setlength{\belowdisplayskip}{3pt}
		L_\text{HB} = \left\| |\bm{S}^{\text{HB}'}| - |\dot{\bm{S}}^{\text{HB}}| \right\|^2 + \left\|\log|\bm{S}^{\text{HB}'}| - \log|\dot{\bm{S}}^{\text{HB}}|\right\|^2
	\end{equation}
	\begin{equation}
		\setlength{\abovedisplayskip}{1pt}
		\setlength{\belowdisplayskip}{2pt}
		L_\text{focal} = -\alpha(1-P_{t,f})^\beta \cdot \log{P_{t,f}}
	\end{equation}
	\begin{equation}
		\setlength{\abovedisplayskip}{1pt}
		\setlength{\belowdisplayskip}{2pt}
		\left\{
		\begin{array}{ccl}
			\bm{S}^{\text{WB}'}_\text{C}&= |\bm{S}^{\text{WB}'}| \odot (|\bm{S}^{\text{WB}'}|+1)^{\frac{\bm{\gamma}-1}{2}} \odot e^{j\bm{S}^{\text{WB}'}_{\text{phase}}} \\
			\dot{\bm{S}}^{\text{WB}}_\text{C}&= |\dot{\bm{S}}^{\text{WB}}| \odot (|\dot{\bm{S}}^{\text{WB}}|+1)^{\frac{\bm{\gamma}-1}{2}} \odot e^{j\dot{\bm{S}}^{\text{WB}}_{\text{phase}}} \\
			\bm{S}_{\text{t}}&= (<\bm{S}^{\text{WB}'}_\text{C},\dot{\bm{S}}^{\text{WB}}_\text{C}> \cdot  \dot{\bm{S}}^{\text{WB}}_\text{C})/||\dot{\bm{S}}^{\text{WB}}_\text{C}||^2 \\
			L_\text{APC}^{\bm{S}^{\text{WB}'}}&=10\log_{10}{\left(||\bm{S}_{\text{t}}||^2/||\bm{S}^{\text{WB}'}_\text{C} -\bm{S}_{\text{t}}||^2\right)} \\
		\end{array} \right.     
		\label{stft_snr}
	\end{equation}
	where $L_\text{APC}^{\bm{S}^{\text{WB}'}}$ and $L_\text{APC}^{\bm{S}^{\text{WB}''}}$ are the loss of CEM and GHCM respectively, and they are calculated as Eq.~(\ref{stft_snr}). $||\cdot||$ denotes the 2-norm of the vector. We set $\alpha\text{=}1$ and $\beta\text{=}2$ in experiment.

	\section{Experiments}
	\subsection{Dataset}
	To evaluate the performance of the updated WB part of HGCN+, we generate 100 hours of WB data with signal-to-noise ratios (SNR) ranging in 0\textasciitilde40~dB using the speech and noise provided by INTERSPEECH 2020 DNS Challenge (DNS-2020) \cite{dns2020}. We divide the data into training and validation set by 4:1. For testing, we generate 540 noisy-clean pairs with 3 SNRs (-5~dB, 0~dB, 5~dB) using the noise and speech that didn't appear in the training or validation set.
	
	For the ICASSP 2022 DNS Challenge (DNS-2022) \cite{dns2022}, we generate a FB set with 3500 hours duration using provided 181 hours of noise, read-English, Russian, French, and Spanish data, and the SNR ranges -5\textasciitilde25~dB. In addition, half of the utterances are convolved with random synthetic and real room impulse responses (RIRs) before being mixed. We divide the data into training and validation by 4:1.

	\subsection{Training setup and comparison methods}
	For the WB experiments, we use HGCN (CEM+GHCM+HM) as the reference and introduce the improvements of each module step by step for comparison ($\text{CEM}^+$, $\text{GHCM}^+$, $\text{HM}^+$, $^+$ denotes the updated version). The $32$~ms hanning window with $25\%$ overlap and 512-point STFT is used. The kernel size is $(5,2)$. The channel number and stride of encoder and decoder are $\left\{12,24,48,64,96,96\right\}$ and $(2,1)$. For HGCN, a 512-units FC after 3-layer 128-units LSTM is adopted. The channel number and stride of GHCM are $\left\{8,16,8\right\}$ and $(1,1)$ and the out channel of last decoder is 22 ($\text{C}_\text{A}=\text{C}_\text{B}=10$). Since the $\text{GHCM}^+$ and the DPRNN are residual, the hidden dimension is the same as the input. The optimizer is Adam \cite{adam}. And the initial learning rate is $0.001$, which decays $50\%$ when the valid loss plateaus for $5$ epochs, and training is stopped if loss plateaus for 20 epochs. 
	
	For HGCN+ in DNS-2022, the $32$~ms Hanning window with $25\%$ overlap and 1536-point STFT is used. The kernel size and stride are $(5,2)$ and $(2,1)$. The channel numbers of the encoder and decoder are $\left\{12,24,48,64,96,96\right\}$ and the channel number of last decoder is 6 ($\text{C}_\text{A}=4$). The hidden cell of the HBM is 256 and 2-layer GRU is used. HGCN+ is trained for 24 epochs on 3500 hours of data with a learning rate of 0.000125. The audio speed is adjusted in 0.9\textasciitilde 1.1 during training.  The number of parameter (Para.) is 5.29~M.

	\subsection{Experimental results and discussion}
	The Real-Time-Factor (RTF) of WB and FB models both are tested on a machine with an Intel(R) Core(TM) i5-6200U CPU@2.30 GHz in a single thread (implemented by ONNX). 
		
	\begin{table}[htp]
		\setlength\tabcolsep{1.4pt}
		\vspace{-0.65cm}
		\centering
		\caption{System comparison on the wide-band test set.}
		\small
		\begin{tabular}{l|c|c|cccc|cccc}
			\toprule
			\multirow{2}*{Model} & \multirow{2}*{\shortstack{Para.\\(M)}} & \multirow{2}*{RTF} & \multicolumn{4}{|c|}{PESQ-WB}& \multicolumn{4}{|c}{STOI(\%)} \\
			\cline{4-11}
			& & & -5dB  & 0dB & 5dB & AVG  & -5dB  & 0dB & 5dB & AVG   \\
			\midrule
			Noisy  & - & - & 1.11 & 1.20 & 1.37 & 1.23  & 72.6 & 81.2 & 88.2 & 80.7  \\	
			HGCN   & \textbf{0.93}  & \textbf{0.11} & 1.59 & 1.95 & 2.37 & 1.97 & 80.2 & 88.2 & 93.6 & 87.4  \\
			\ +$\text{CEM}^+$ & 3.60 & 0.17 & 1.62  & 2.00 & 2.44 & 2.02 & 82.8 & 90.3 & 94.5 & 89.1    \\
			\ \ +$\text{GHCM}^+$ & 4.12 & 0.16 & 1.62  & 2.01 & 2.47 & 2.03  & 83.0 & 90.3 & 94.5 & 89.2   \\
			\ \ \ +$\text{HM}^+$ & 4.11 & 0.14 & \textbf{1.65}  & \textbf{2.04} & \textbf{2.48} & \textbf{2.06} & \textbf{83.6} & \textbf{90.8} & \textbf{94.8} & \textbf{89.7}  \\
			\bottomrule
		\end{tabular}
		\label{tab:1}
		\vspace{-0.4cm}
	\end{table}
	To evaluate the performance of WB models (100 hours training data), two metrics are utilized, namely PESQ (PESQ-WB, PESQ-NB (narrow-band, 0\textasciitilde 4~KHz)) and STOI \cite{pesq,stoi}. The comparison based on WB set is shown in Table~\ref{tab:1}. It can be seen that the dual-path encoder and DPRNN in $\text{CEM}^+$ bring an improvement in terms of feature utilization and frequency dependencies modeling with more arithmetic complexity. $\text{GHCM}^+$ improves the performance because the linear module has a wider receptive field with less computational complexity than the convolution. In $\text{HM}^+$, the significance-based VRD reduces the computation of CEM, and the introduction of cosine makes gates more accurate and instructive. 
	
	\begin{table}[htp]
		\centering
		\setlength\tabcolsep{3pt}
		\vspace{-0.6cm}
		\caption{System comparison on DNS-2020 synthetic test set.}
		\begin{tabular}{lcccc}
			\toprule
			Model & Para.(M) & PESQ-WB & PESQ-NB & STOI(\%)\\
			\midrule
			Noisy & -  & 1.58 & 2.45 & 91.5 \\	
			DCCRN \cite{dccrn}  & \textbf{3.67} & - & 3.27 & -    \\
			GaGNet \cite{gaze} & 5.94 & 3.17 & 3.56 & 97.1  \\
			HGCN+ & 5.29 & \textbf{3.19} & \textbf{3.65} & \textbf{97.2}  \\
			\bottomrule
		\end{tabular}
		\label{tab:dns2020}
		\vspace{-0.3cm}
	\end{table}

	\begin{table}[htp]
		\vspace{-0.618cm}
		\setlength\tabcolsep{1.5pt}
		\centering
		\caption{Challenge results for track 1 on DNS-2022.}
		\begin{tabular}{lcccccc}
			\toprule
			Model  & Para.(M) & SIG & BAK & OVRL & WAcc & Final Score\\
			\midrule
			NSNet2\cite{nsnet} & 6.17   & 3.62 & 3.93 & 3.26 & 0.63 & 0.60 \\
			HGCN+ & \textbf{5.29} &  \textbf{4.01} & \textbf{4.55} & \textbf{3.81} & \textbf{0.65} & \textbf{0.68}\\
			\bottomrule
		\end{tabular}
		\vspace{-0.3cm}
		\label{tab:dns2022}
	\end{table}
	
	We evaluate the HGCN+ (3500 hours training data) on the DNS-2020 and the DNS-2022 test sets. Because HGCN+ processes the HB and WB spectra separately, and the WB spectrum contains more speech information, the model has good performance on the WB test set can bring gains to the processing of FB signal, as shown in Table~\ref{tab:dns2020}. In DNS-2022, ITU-T P.835 framework \cite{p835} and Word Accuracy (WAcc) are used to evaluate the speech quality, and HGCN+ outperforms the baseline \cite{nsnet} with less parameters in all metrics, as shown in Table~\ref{tab:dns2022}. By audiometric analysis, we also found that non-speech sound, such as breath, sob, etc. is usually regarded as noise and filtered out, which results in incomplete human voice but better speech and the BAK MOS which is relatively larger than SIG MOS. In addition, our method makes a confusing pitch choice in the case of multi-speaker and degrades the performance also. The RTF of HGCN+ in processing the FB signal is 0.16 and it consumes 5.22~ms per frame. The frame size and overlap are 32~ms and 25\%. Since the model is causal without looking forward, the latency is $32\text{+}8\text{=}40$~ms, which satisfies the requirements of the challenge. 

	\section{Conclusion}
	In this paper, we improve each module of our HGCN. First, we model the harmonic integration by cosine and propose a significance-based VRD to predict the harmonic locations efficiently. Second, we introduce the dual-path encoder, DPRNN, and residual linear structure to CEM and GHCM to enhance model performance. Finally, we add a high-band module to help the model handle the FB signal, resulting in HGCN+. HGCN+ outperforms the referenced models on the DNS-2020 and DNS-2022 test sets.
	
	\bibliographystyle{IEEEbib}
	\bibliography{strings,refs}
	
\end{document}